\documentclass[twocolumn,showpacs,preprintnumbers,amsmath,amssymb,aps,prb]{revtex4-1}

\usepackage{graphicx}

\usepackage{dcolumn}

\usepackage{bm}

\usepackage{color}

\begin{document}

\title{Controlling conductance statistics of quantum wires by driving ac fields }

\author{V\'ictor A. Gopar}

\affiliation{Depto de F\'isica Te\'orica, Facultad de Ciencias, and Instituto de Biocomputaci\'on y
F\'isica de Sistemas Complejos (BIFI), Universidad de Zaragoza, Pedro Cerbuna 12, E-50009, Zaragoza, Spain.}

\author{Rafael A. Molina}

\affiliation{Instituto de Estructura de la Materia, CSIC, Serrano 123, 28006 Madrid, Spain.}

\begin{abstract}

We calculate the entire distribution of the conductance $P(G)$ of a one-dimensional disordered system --quantum wire-- subject to a time-dependent field. Our calculations are based on Floquet theory and a scaling approach to localization.
Effects of the applied ac field on the conductance statistics can be strong and in some cases dramatic, as in the high-frequency
regime where the conductance distribution shows a sharp cut-off. In this frequency regime, the conductance is written as a product of a frequency-dependent term and a field-independent term, the latter containing the information on the disorder in the wire.
We thus use the solution of the Mel'nikov equation for time-independent transport to calculate $P(G)$ at any degree of disorder. At lower frequencies, it is found that the conductance distribution and the correlations of the transmission Floquet modes
are described by a solution of the Dorokhov-Mello-Pereyra-Kumar equation with an effective number of channels.
In the regime of strong localization, induced by the disorder or  the ac field, $P(G)$ is a log-normal distribution. Our theoretical results are verified numerically using a single-band Anderson Hamiltonian.

\end{abstract}


\maketitle

\section{Introduction}

Quantum electronic transport in disordered structures has been of fundamental and practical interest since seminal ideas by Anderson and Mott.\cite{AndersonMott}
Recently, transport in driven systems has received much attention since new  phenomena have been found and the application of a time-dependent driving field opens more possibilities to control the transport in an electronic device. For instance, it has been proposed the possibility of manipulating the electronic transport through molecular structures by applying  an ac field. \cite{Lehmann2002, Kohler2004,Rey2005}

Disorder in a sample can be an unavoidable and unwanted ingredient in transport experiments, however, it  can also be seen as an ingredient to be exploited in order to manipulate the transport properties of a system. The disorder gives a random character to the transport and a statistical analysis naturally emerges.
In the absence of ac fields, the statistics of quantities such as the conductance has been widely studied. At present, there is a good understanding of the effects of disorder on the statistical properties of the conductance for one- and quasi-one dimensional systems within a non-interacting electron model. In fact, the distribution of conductances is known, within a scaling approach to localization, at zero temperature and infinitesimally small applied voltage: \cite{pier-book,carlo-review} for one- and quasi-one dimensional systems the evolution of the conductance distribution as a function of the length of the sample is described by the Mel'nikov and Dorokhov-Mello-Pereyra-Kumar (DMPK) equations, respectively. For finite temperatures and bias voltages as well as higher dimensions, some progress has been made on the description of the conductance statistics.\cite{mu-go}

In contrast to the well studied problem of electronic transport in nondriven disordered systems and in spite of the possible interest for applications, remarkably little is known about the statistical properties of transport quantities when a time-dependent field is applied; effects of energy fluctuations on the current statistics in short molecular wires has been studied in Ref. [\onlinecite{Kaiser}], while effects of ac fields on the localization length and the conductance distribution of a  ring driven by a time-dependent magnetic flux have been studied in Refs. [\onlinecite{Martinez06a}] and [\onlinecite{lili}], respectively.

\begin{figure}
\includegraphics[width=0.9\columnwidth]{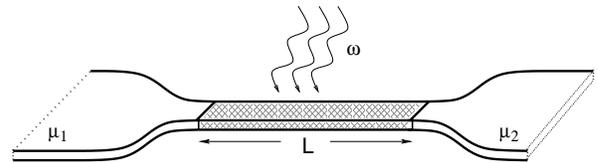}
\caption{Schematic view of a one-disordered system of length $L$ driven by a frequency-dependent force. The quantum wire is connected to electron reservoirs $\mu_1$ and $\mu_2$.}
\label{fig_1}
\end{figure}

In this work, we calculate the complete distribution of the conductance of a disordered quantum wire subject to a time-periodic driving field. We obtain the distribution from high to low frequencies with  different regimes of disorder strength and localization. Our theoretical results are compared to tight-binding numerical simulations.

\section{Tight-binding model and Floquet-Green Theory}

We start by describing briefly the model of our driven quantum wire. The conductance
is calculated adopting a Floquet scattering approach to the electronic transport problem generalizing the Landauer-B\"uttiker formulation to driven systems.  \cite{Kohler,Buttiker}

We might think of an experimental setup where a disordered sample, attached by perfect leads to a reservoir of electrons on each side, is subject to a laser beam with an angle of incidence perpendicular to the wire and a polarization angle parallel to the wire, Fig. 1. We assume that such a system is described by the following time-dependent Hamiltonian
\begin{equation}
\label{totalhamiltonian}
 H(t)= H_{\mathrm{w}}(t)+H_{\mathrm{l}}+H_{\mathrm{c}} .
\end{equation}
The wire Hamiltonian $H_{\mathrm{w}}(t)$  is time-periodic: $H_{\mathrm{w}}(t)=H_{\mathrm{w}}(t+T)$, where $T=2\pi/\omega$, $\omega$ being the frequency of
the applied field. For each term in Eq. (\ref{totalhamiltonian}) we have:
\begin{eqnarray}
H_{\mathrm{w}}(t) &= &
 -\frac{\Delta}{4}\sum_{j=1}^{N-1} \left(|j\rangle\langle j+1|+|j+1\rangle\langle j| \right) \nonumber   \\
 && + 2V\cos{(\omega t)}\sum_{j=1}^N |j\rangle j \langle j| ,
\end{eqnarray}
where $N$ is the total number of sites of the wire, $\Delta$ is the band width of the nondriven system, and $V$ is the field amplitude (a factor of 2 is introduced by computational convenience). The Hamiltonian for the leads
$H_{\mathrm{l}}$ is modeled as
\begin{equation}
 H_\mathrm{l} = \sum_k \epsilon_k\left(  c^{\dagger}_{Lk} c_{Lk}  +   c^{\dagger}_{Rk} c_{Rk}  \right) ,
\end{equation}
where $c^\dagger_{L(R)k}$ is the creation operator for an electron at the left lead (right) with momentum $k$ and
the  coupling lead-wire Hamiltonian $H_{\mathrm{c}}$ is given by
\begin{equation}
 H_{\mathrm{c}}=\sum_ k V_{1k}c_{Lk}^{\dagger}c_1+V_{Nk}c^{\dagger}_{Rk}c_N + \mathrm{ H. c.}
\end{equation}
The coupling can be described by the spectral density: $\Gamma_{N(1)}(E)=2\pi\sum_k|V_{R(L)k}|^2\delta(E-\epsilon_k)$.
The driving field might have an effect on the coupling leads-wire, however, this
situation can be mapped onto a Hamiltonian of the form given by Eq. (\ref{totalhamiltonian}) by a gauge transformation. \cite{Kohler} We use the wide-band approximation where we assume that the coupling between leads and wire is energy independent. This treatment is justified whenever the conduction bandwidth of the leads is much larger than all other relevant energy scales in our problem. We use it as we are interested in the effect of the driving in the conductance distribution and not of the details of the coupling to the leads.
In this approximation the information of the wire-coupling can be introduced via two self-energies at the first and last sites of the wire
($\Gamma_1$ and $\Gamma_N$) transforming the wire Hamiltonian to a non-Hermitian Hamiltonian. Under the above assumptions
and within a tight binding approximation, the Hamiltonian associated to our quantum wire can be written as
\begin{eqnarray}
& &H(t)=-\frac{\Delta}{4}\sum_{j=1}^{N-1} \left(|j\rangle\langle j+1|+|j+1\rangle\langle j| \right)+ \sum_{j=1}^N E_j|j\rangle \langle j| \nonumber \\
& &+2V\cos{(\omega t)}\sum_{j=1}^N |j\rangle j \langle j| +i\Gamma_1 |1\rangle  \langle 1| +i\Gamma_N |N\rangle \langle N| ,
\label{tightH}
\end{eqnarray}
where $E_j$ is the random on-site energy, distributed uniformly in the interval $(-W/2, W/2)$. The form of the external time-dependent field assumes a dipolar approximation for a monochromatic field so the wavelength of the laser light must be longer than the length of the wire. As the Hamiltonian is time-periodic, the Floquet theorem states that
there is a set of solutions $|\phi_\epsilon^{\alpha, m}(t)\rangle$--Floquet states--to the equation
\begin{equation}
\label{floquetequation}
 \left[H(t)-i\hbar \frac{d}{dt}\right]|\phi_\epsilon^{\alpha, m}(t)\rangle =\epsilon^{\alpha, m}|\phi_\epsilon^{\alpha, m}(t)\rangle ,
\end{equation}
where  $\epsilon^{\alpha, m}=\epsilon_\alpha+m\hbar\omega$ with $m$ an integer number and $-\hbar\omega/2\le {\mathrm Re}(\epsilon_\alpha) \le \hbar \omega/2$.
The Green's function $ G(E,t',t'')$ satisfying $\left[ {\mathbb I} E - H(t') \right]G(E,t',t''){\mathbb I}={\mathbb I}\delta_T(t'-t'')$, where ${\mathbb I}$ is the identity operator and $\delta_T(t)$ is a $T$-periodic delta function, can be written in terms of the Floquet
states $|\phi_\epsilon^{\alpha, 0}(t)\rangle$ . Thus the Fourier components $G^k(E)$ of $G(E)$ can be written as
\begin{equation}
\label{Gk}
 G^{(k)}(E)=\sum_{\alpha,m}\frac{|\phi_{k+m}^{\alpha, 0}\rangle \langle {\phi_m^{\alpha, 0}}^\dagger|}{{E-\epsilon_\alpha-m\hbar\omega}} .
\end{equation}
We define the dc-conductance for an ac-driven quantum wire as
\begin{equation}
G=\lim_{V \rightarrow 0} \frac{d \bar{I}}{dV} ,
\end{equation}
where $\bar{I}$ is the current averaged over one period of the driving field.
Thus $G$ is an experimentally accessible quantity which can be written as a sum of the Floquet modes
$g^{(k)}(E_F)$:
\begin{equation}
\label{G}
 G=\sum_{k=-\infty}^\infty g^{(k)}(E_F) ,
\end{equation}
where
\begin{equation}
\label{gk}
g^{(k)}(E_F)= \frac{1}{2} \left[ T_{1N}^{(k)}(E_F) + T_{N1}^{(k)}(E_F) \right] ,
\end{equation}
with $T_{1N}^{(k)}(E)=\Gamma_1\Gamma_N|G_{1N}^{(k)}(E)|^2$ being the transmission for electrons from the left to the right lead, similarly   $T_{N1}^{(k)}(E)=\Gamma_N\Gamma_1|G_{N1}^{(k)}(E)|^2$ for electrons from the right to the left. In the static case, $T^{(k)}_{1N} = T^{(k)}_{N1}$
 for systems with time-reversal symmetry, however, in the presence of time-dependent fields this is not longer true \cite{Kohler}. When both transmissions are not equal, a current at zero voltage can be induced
(pumped current). This pumped current appears as an offset at zero voltage in the $I\mathrm{-}V$ characteristic of our system. The presence of a pumped current does not modify our formulas, as we calculate the slope of the $I\mathrm{-}V$ curve at zero
voltage. In the systems we considered the pumped current is very much reduced as the length of the systems
increases and is, in general, negligible, in contrast to results by Kaiser {\it et. al.} in Ref. [\onlinecite{Kaiser}] ,
where they study the current in shorter molecular wires ($N=6$).

We are thus interested in the statistics of the conductance $G$ given by Eq. (\ref{G}). Our theoretical predictions are verified numerically by sampling over different disorder realizations, i.e., over different configurations of $E_j$ in Eq. (\ref{tightH}). The numerical simulations are performed according to Eqs. (\ref{Gk}) and (\ref{G}), where the Fourier components $G^{(k)}$ are calculated using the method of matrix continued fractions. \cite{Martinez03,Martinez06a,Martinez08}  We assume perfect lead-wire coupling  ($\Gamma_1=\Gamma_N=1$) although different coupling strengths can be implemented in our numerical simulations and theoretical framework. The histograms shown in all the figures of this work were
obtained from 2000 different disorder realizations.


\section{Results}

\subsection{High-frequency regime}
\begin{figure}
\includegraphics[width=\columnwidth]{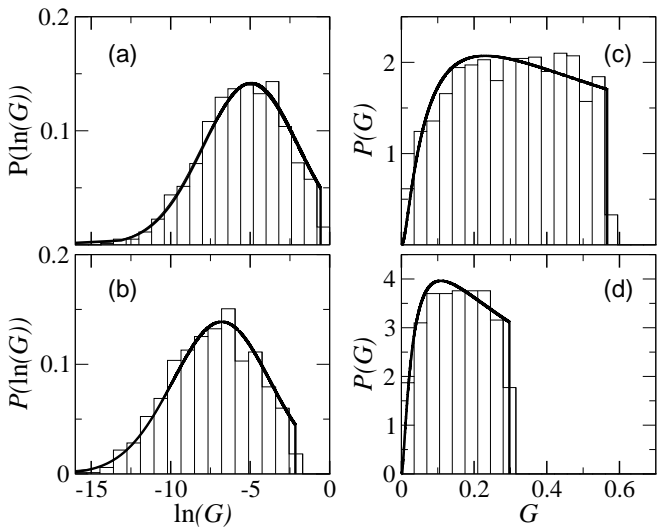}
\caption{Conductance distributions $P[\ln (G)]$ for strength disorder $W=5$, field parameters $V=4$ and (a)  $\omega=35$  and (b) $\omega=20$. $L/l=4.95$ and 5.20 for (a) and (b), respectively. $P(G)$ for strength disorder $W=2$ and field parameters $V=4$, (c)  $\omega=35$ and (d) $\omega=25$. $L/l=0.8$ and 0.83 for (c) and (d), respectively. A good agreement between theory (solid lines) and numerical simulations (histograms) is seen.}
\label{fig_2}
\end{figure}
We start with the simplest case which turns out to be the high frequency regime characterized by frequencies  $\hbar \omega > (\Delta + W)/2$. It is also convenient to begin with this regime since it illustrates in a simple way the approach to our conductance calculations along this work.

In the absence of disorder and for high frequencies
it has been shown that the conductance is proportional to the square of
the Bessel function $J_0(2VL/\hbar\omega)$. \cite{Martinez08}
A key point is that this frequency-dependent factor $J_0(2VL/\hbar\omega)$ remains under the presence of disorder. \cite{Martinez06a}
Thus, for a disordered wire we write the conductance $G$ as the product
\begin{equation}
\label{Gproptog}
G = J_0^2(\nu) g ,
\end{equation}
where $\nu=2VL/\hbar\omega$ and $g$ is the dimensionless conductance of the disordered wire in the absence of the driving field.  The calculation of the  distribution of $G$ is now straightforward since the conductance distribution
for the static case $p(g)$ is given by the solution of the Mel'nikov equation, which is given by \cite{carlo-review}
\begin{equation}
\label{pofg}
p(g)=\frac{1}{\sqrt{2\pi}}\Big(\frac{1}{s}\Big)^{\frac{3}{2}}
\frac{{\rm e}^{-s/4}}{g^2}\int_{y_0}^{\infty}dy\frac{y{\rm e}^{-y^2/4s}}
{\sqrt{\cosh{y}+1-2/g}},
\end{equation}
where $y_0={\rm arccosh}{(2/g-1)}$ and $s=L/l$, l being the mean free path. $s$ can be seen as
a disorder parameter which can be obtained from the disorder-average: $s=\langle -\ln g \rangle$. In order to calculate $P(G)$ we just need to make the change of variable $g \to G$ in Eq. (\ref{pofg}), using Eq. (\ref{Gproptog}). We can work with the exact expression  (\ref{pofg}), however, we provide an analytical expression, which can be obtained by the saddle-point method, for $P(G)$:
\begin{eqnarray}
\label{pofGapprox}
&& P(G)=\frac{C_G}{G^{3/2}}\frac{J_0^{3/2}(\nu)}{(J_0^2(\nu)-G)^{1/4}} \sqrt{\mathrm{acosh}{\left({J_0(\nu)}/{\sqrt{G}}\right)}} \nonumber \\ &\times&
\exp{\left[ -({l}/{L})\mathrm{acosh}^2\left({J_0(\nu)}/{\sqrt{G}}\right) \right]} ,
\end{eqnarray}
where $C_G$ is a normalization constant and the ratio $L/l$ is obtained from Eq. (\ref{Gproptog}) as
\begin{eqnarray}
\label{newlocalization}
{L}/{l} &=&2  \ln J_0(\nu)-\langle \ln G \rangle .
\end{eqnarray}
We notice that the $L/l$ is the only parameter in Eq. (\ref{pofGapprox}) which can be extracted from the numerical experiments using Eq. (\ref{newlocalization}); in this sense, we have a free parameter theoretical result.
In Fig. \ref{fig_2} we compare Eq. (\ref{pofGapprox})  with the numerical simulations for different values of disorder, frequencies, and field amplitudes. A good  agreement is seen in all cases. We point out the sharp cut-off of the distributions, which is determined by the frequency-dependent term $J_0(\nu)$ in Eq. (\ref{Gproptog}).
This is a manifestation of the phenomenon of coherent destruction of tunneling or dynamical localization
by the interference of multiple paths due to the rapid oscillations of the external field. \cite{CDT,mu-wo-ga-go} This strong effect on the statistics suggests the possibility of controlling the conductance: by tuning properly the applied field the conductance through the disorder sample can be completely switched off at a desired conductance value.  We finally remark that the complete conductance
distribution is determined by the frequency-dependent term $J_0(\nu)$  and the disorder parameter  $L/l$ in Eq.  (\ref{newlocalization})


\subsection{Intermediate and low-frequency regimes}

Let us consider first the strong localization regime ($L >> l$) where the analysis of the conductance statistics is particularly simple. As we shall see, when $L$ becomes of the same order of $l$ the calculations of the distribution of the conductance are more involved.

When we increase the disorder strength and/or choose the field parameters (amplitude, frequency) such that
the system is strongly localized, we observe that the Floquet modes are strongly correlated; we show two typical examples of this behavior in the insets of Fig. 3, where we plot the distribution $F(\delta g)$ of difference between the first and second transmission Floquet modes  $\delta g$ (we have divided this difference between modes by  $\langle G \rangle$ with the only purpose of measuring $\delta g$ in units of conductance average). We  observe that $F(\delta g)$ is highly concentrated at small values of $\delta g$. We point out  that  in the localized regime $G << 1$ and since $G$ is the sum of the Floquet modes,  Eq. (\ref{G}), it is expected that each Floquet mode is in the localized regime. On the other hand, we recall that in the static case, the conductance distribution follows a log-normal distribution. Thus, the distribution of the logarithm of the total conductance is given by the convolution of log-normal distributions, i.e.,  each of these distributions corresponds to a Floquet mode. The resulting distribution is again a log-normal distribution. Therefore we write $P(\ln G)$ as
\begin{equation}
\label{poflogG}
 P(\ln G)= C_G \exp{\left[- (\ln G-\langle \ln G \rangle)^2/2\sigma^2{(\ln G)}\right]} .
\end{equation}
\begin{figure}
\includegraphics[width=\columnwidth]{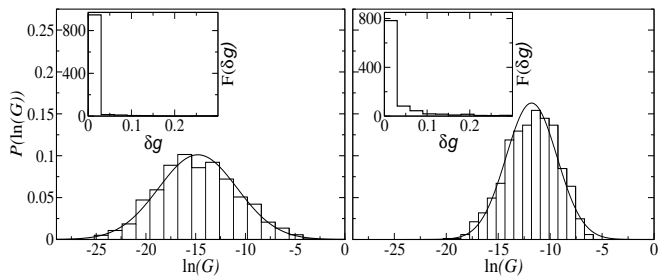}
\caption{Insets: distribution of $\delta g=|g0-g1|/\langle G \rangle$ for $L=10, W=5, V=4$,  $\omega=2$, and $\langle G \rangle=6.7\times 10^{-5}$ (left); and $L=10, W=2, V=4, \omega=2$ with $\langle G \rangle=1.7 \times 10^{-4}$ (right). Main frames: $P[\ln (G)]$ for the same parameters of the insets. Solid lines are given by Eq. (\ref{poflogG}) with $\langle \ln G \rangle= -14.7 $ and $\sigma^2(\ln G)= 15.5 $ (left) extracted from the numerical simulation. $\langle \ln G \rangle=-11.7 $ and $\sigma^2(\ln G)=6 $ for the right panel.}
\label{fig_3}
\end{figure}
In Fig. 3 (main frames) we show $P(\ln G)$ for two different disorder strengths. The average and variance of  $\ln G$ are extracted from the numerical data and substituted into Eq. (\ref{poflogG}). A good agreement between theory and numerics is seen.

Perhaps the most interesting case is when the system is not strongly localized by the ac field and/or  disorder.
We go further with our approach to the problem where the statistics of the frequency-dependent conductance fluctuations are derived from the statistical properties of the transport problem of the static case. For instance,
we have found that the Floquet modes $g^{(k)}$ show similar correlations to the conductance channels of the multichannel (static) case. For simplicity we shall consider the case where
two Floquet modes $g^{(0)}$ and $g^{(1)}$ ($g^{(-1)}$ is also considered, statistically, it gives the same contribution as $g^{(1)}$) give the main contribution to  $G$. \cite{modes}

We recall that the DMPK equation is derived for a disordered wire whose length $L$ is much larger than
its width $L_y$  ($L_y << L $), i.e., a quasi-one-dimensional system. Within this limit the diffusion of electrons in the transverse direction can be neglected, although several
transverse modes or channels might be open and contribute to the conductance.  For an energy of electrons $\mathcal E$ and for a given channel $n$, the longitudinal energy ${\mathcal E}_n$ is given by ${\mathcal E}_n={\mathcal E}- e_n$, where $e_n=n\pi/L_y$, $n$ being a positive integer.  It is important to remark that within the DMPK framework only longitudinal
diffusion is considered and  the information of the number of modes is given  through the energies of the finite number of open channels.
In our one-dimensional wire driven by an periodic ac field, the eigenvalues of the Schr\"odinger equation are given by
$\epsilon^{\alpha, m} = \epsilon_\alpha+m\hbar w$, an expression similar to the static case; however, $m$ takes unrestricted values (positive and negative) whereas the open channels $n$ are finite in the static case.
With this argumentation we expect that the statistical properties of the time-dependent problem can
be described through the known statistical properties of the static case for a quasi-one-dimensional geometry.  As we have mentioned above, we  consider the case where two Floquet modes give the major contribution to the conductance. We shall  thus
use the DMPK results  for two open channels. Within the DMPK framework,
the joint distribution of the variables $x_n$ related to the conductance $g_n$ by
$x_n=\mathrm{arccosh}{\sqrt{1/g_n}}$, where $n$ labels the channels ($n=1,2$), is given by\cite{rejaei}
\begin{equation}
\label{pofx1x2}
 p(x_1,x_2)= \exp[{u(x_1,x_2)+V(x_1)+V(x_2)}] ,
\end{equation}
where $u(x_1,x_2)=\ln {[|\sinh^2 x_1 -\sinh^2 x_2|| x^2_1 - x^2_2|^2]^{1/2}}$,  $V(x_n)=-3x^2_n/2s+\ln{|x_n^2\sinh 2x_n|^{1/2}}$. Equation  (\ref{pofx1x2}) is an approximation to the solution of the DMPK equation and it is useful from strong to weak localized regimes \cite{mu-wo-go}.
\begin{figure}
\includegraphics[width=\columnwidth]{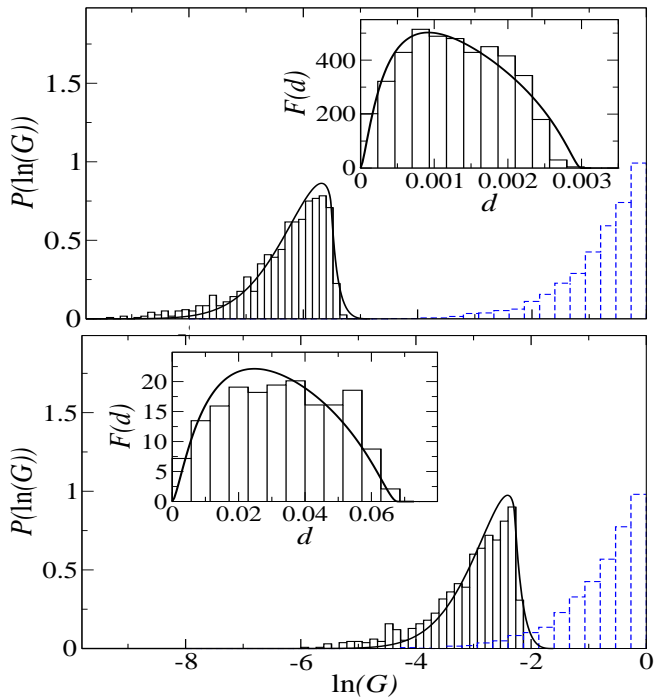}
\caption{Main frames: conductance distributions for $L=10, W=2, V=1.0$, $\omega=4$,  $a=0.003$ (top) and $\omega=5$,  $a=0.068$ (bottom). The solid line is the analytic result with $s=1.5$ (top) and $s=1.35$ (bottom). The histograms in broken lines correspond to the numerical results when the field is switched off. Insets: distribution of $d=g^{(0)}-[g^{(1)}+g^{(-1)}]$ for the same parameters of the main frames; theory (solid lines) and numerical calculations (histograms) show  a good agreement. }
\label{fig_4}
\end{figure}
As we have discussed above, when the field is applied, the Floquet modes follows a statistics given by Eq. (\ref{pofx1x2}); in order to illustrate this fact we calculate numerically the distribution of the Floquet modes difference  $d=g^{(0)}-[g^{(+1)}+g^{(-1)}]$, $F(d)$, and compare to the theoretical distribution calculated from Eq. (\ref{pofx1x2}). The disorder parameter $s(=L/l)$ in Eq. (\ref{pofx1x2}) is fixed by fitting the average $\langle d \rangle$ to its  corresponding numerical result.
In the insets of Fig. 4, $F(d)$ is plotted for two different frequency values. As we can see the trends of the numerical histograms
are well described by the theoretical $F(d)$ (solid line). These results for $F(d)$ give us  confidence in using the solution of the DMPK equation, Eq. (\ref{pofx1x2}).  We then calculate  $G$ assuming that it
can be expressed as the product of a frequency-dependent part, $a$, and a static contribution, $g$: $G=a g$, i.e., the frequency-dependent term works as a rescaling factor to the static conductance, in a similar manner to $J_0(\nu)$  in the high-frequency regime studied previously. This factor is estimated as the ratio $a= \langle G \rangle/\langle g \rangle$, where $\langle g \rangle$ is the conductance average when the ac field is switched off. Thus, the distribution  $P(G)$ is given by
\begin{equation}
\label{pofGinter}
 P(G)=\left\langle \delta \left[{G}/{a}-\left({1}/{\cosh^2x_1}+{1}/{\cosh^2 x_2}\right)\right]\right\rangle ,
\end{equation}
where $\langle \cdot \rangle$ denotes average performed accordingly to Eq. (\ref{pofx1x2}).
In Fig. 4 (main frames) we compare the theoretical distributions (solid line) as given by Eq. (\ref{pofGinter}) and the numerical results (histograms) for two different frequencies: a good agreement is seen in both cases.
For distinguishing details of the distributions, we plot $P(\ln G)$ instead of $P(G)$. We also plot $P(\ln G)$
for the static case (histograms in broken line) in order to remark the strong effect of the ac field.  Notice the
sharp decay of the distributions, although it is not as strong as the high frequency regime. The distributions are also shifted  from the origin due to renormalization value of the conductance $a$ in Eq. (\ref{pofGinter}).  The value of  $a$ depends on some complex interference pattern and as the results in  Fig. 4 show,  it can vary widely with a small variation
in the parameters of the model.

\subsection{Summary and conclusions}

The statistical properties of the quantum transport in driven disordered systems by ac fields
are not well known. Here we have studied the  effects of an applied ac field on the conductance distribution  under different conditions by varying the disorder strength and field parameters. The statistical properties of the frequency-dependent problem, in particular the conductance distribution, are obtained from known statistical properties of the transport in the static case. We have found of special interest in the regime of  high frequencies  where the conductance distribution shows a sharp cut-off which suggests the possibility of switching applications by tuning the ac field, although, other aspects such as heating or electrons interactions, which we neglect, might be relevant to an accurate description of an experimental realization.
This cut-off is a manifestation of dynamical localization or coherent destruction of tunneling due to multiple interference. Coherent destruction in time-periodic potentials
has been seen in cold atoms and semiconductor supperlatices. \cite{experiments} It should be of interest to
observe this effect in quantum transport experiments.

This work was supported by the Programa Ram\'on y Cajal of MICINN (Spain), the European Social Fund,
and  MICINN (Spain) under  Project No.~FIS2009-07277.


\begin{thebibliography}{50}
\bibitem{modes} We have found numerically that the number of relevant Floquet modes to $G$ is approximately given by $3LV/2\omega$.
\end{thebibliography}
\end{document}